# DESIGN STATUS OF BCC CRYOMODULE FOR LCLS-II HE*

C. Narug, T. Arkan, S. Cheban, M. Chen, B. Hartsell, J. Kaluzny, Y. Orlov, V. Kashikhin
Fermi National Accelerator Laboratory, 60510 Batavia IL, United States

## Abstract

A Buncher or Capture Cavity (BCC) Cryomodule is currently in development at Fermilab for use in a second injector for LCLS-II-HE. The BCC Cryomodule is designed to contain one 1.3 GHz cavity and one solenoid magnet as part of a 100MeV low emittance injector. The design considerations for the Cryomodule are similar to the LCLS-II cryomodule with additional requirements to account for additional vacuum loading at the end of this vessel due to the termination of the insulating vacuum. To accomplish this design, the cryomodule is being developed using the experience gained during the development of the LCLS-II cryomodule. The design, analysis, and status of the Cryomodule will be discussed.

## PROJECT OVERVIEW

A Buncher or Capture Cavity (BCC) Cryomodule is currently in development at Fermilab for use in a second injector for the High Energy upgrade of the Linac Coherent Light Source (LCLS-II HE). The BCC Cryomodule is designed to contain one 1.3 GHz cavity and one solenoid magnet as part of a 100 MeV low emittance injector. The current design of the vessel can be seen in Figure 1 and a cross sectional view can be seen in Figure 2. There are five general regions of the design of the assembly; the cavity, the magnet, internal piping, cryostat shell, and the support stands. The cavity will be the same 1.3 GHz cavity that is used in the LCLS-II HE cryomodules. The magnet that will be used is a combined solenoid, dipole, and quadruple design. The internal piping, cryostat shell, and support stands will all be designed to be compatible and reuse design aspects of the other LCLS-II HE Cryomodules. The BCC Cryomodule is located at the end of the assembly, it will be required to adapt to some additional loading/design requirements. The piping system is required to be terminated or be recirculated at the end of the vessel. Due to the insulating vacuum, a pressure load will be present at the end of the assembly which will require additional strength in the supports and consideration of load paths. The Cryomodule will also need to be supported on a adjustable stand to for alignment.

## SAFTEY REQUIREMENTS

The BCC cryomodule will be installed and be designed to be the saftey requirements of Stanford Linear Accelerator Center [1] [2]. Equipment built and tested at Fermilab are also required to meet the Fermilab Environment, Safety and Health Manual [3]. The saftey requirements for both laboratories require components to be designed using design codes. The internal piping system will be designed to meet the requirements of the American Society of Mechanical Engineers (ASME) B31.9 Piping requirements. The main components of the Cryomodule will be designed to meet the ASME Boiler and Pressure Vessel Code (BPVC) Section VIII Division 1 [5] with additional methods used from the ASME BPVC Section VIII Division 2 [6]. The cryomodule stand is be designed to meet Americian Institute of Steel Construction (AISC) Specification for Structural Steel Buildings [8] while using design by analysis methods to verify components that are outside of the scope of the specification. The cryomodule will also meet the requirements of the American Society of Civil Engineers (ASCE) Minimum Design Loads and Associated Criteria for Buildings and Other Structures [7].

* Work supported by Fermi Research Alliance, CONF-

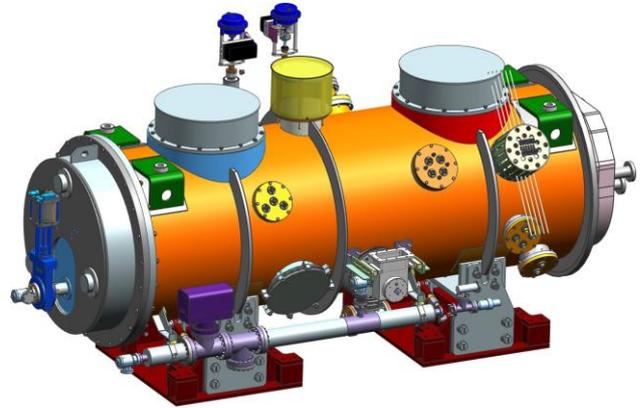

Figure 1: BCC Cryomodule

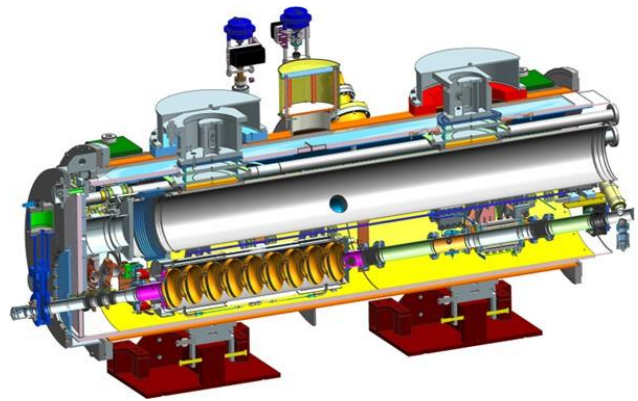

Figure 2: BCC Cryomodule Cross Sectional View

## ANALYSIS PERFORMED

In the design of the BCC Cryomodule four sets of analysis that have been performed; a magnet design, a piping analysis for the cryogenic system, a transportation analysis, and structural analysis for the shell and supporting components.

## MAGNET SYSTEM

The magnet system is comprised of a combined Solenoid, Dipole and Quadruple magnet. The analysis of the magnet was performed to design and optimize the magnet parameters. Parameters for the magnet have been defined to ensure compatibility with other LCLS-II Cryomodules. A series of field simulations were performed to validate the magnets to the BCC magnet system specifications. The results of the field simulations can be seen in Figure 3.

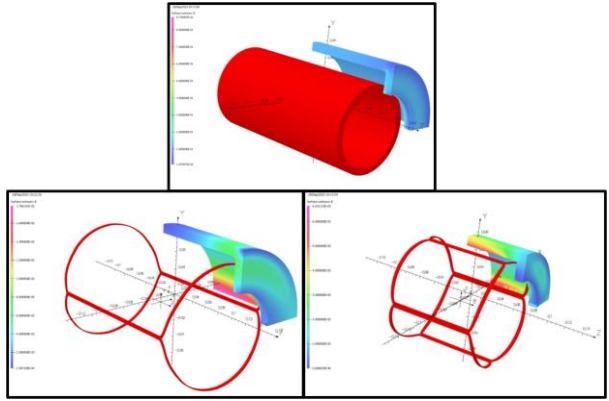

Figure 3: Field Simulations: Solenoid Magnet (Top), Dipole Magnet(Left), Quadrupole Magnet (Right)

## PIPING SYSTEM

A series of cryogenic supply lines are used to control the cooldown and temperature of the components inside the cryomodule during operation. The cryogenic piping lines connect directly to the LCLS-II HE cryomodules and will be terminated in the BCC cryomodule. The lines passing through the vessel can be seen in 4. The pressure lines in the system range on the order of 0.0031 bar.a up to 3.7 bar.a. To ensure the components will not be over-pressurized during warm up, lines A,C,D,E,F,and H have a design pressure of 20 bar.a while the lines B and G will have cold design pressure of 4.1 bar.a. Aspects of the venting scenarios, seismic behavior, and welding processes have been considered and integrated into the design. The piping system for the BCC Cryomodule is designed and analyzed to meet the requirements of ASMSE B31.3 [4].

## TRANSPORTATION ANALYSIS

The cryomodule will be transported using the shipping frame designed for the LCLS-II HE Cryomodules. The shipping frame will be adjusted to account for the shorter length of the cryomodule. The cryomodule will be attached to the frame using a series of damping spring supports to reduce the relative acceleration and resonant frequencies of the cryomodule. The analysis is being preformed to ensure that no critical components have a resonant frequency less than 15 Hz and make any changes that could be needed to ensure damage will not occur during shipment. The cryomodule in the shipping frame can be seen in Figure 5.

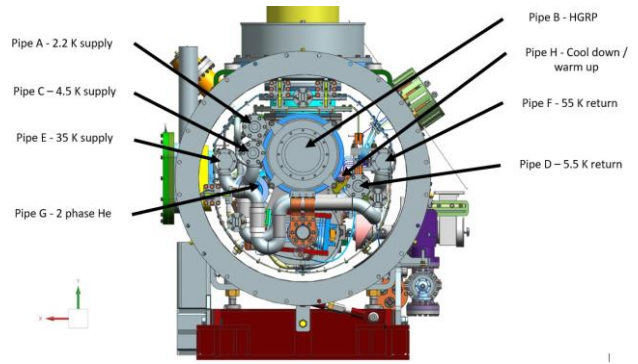

Figure 4: Cryogenic Piping Lines

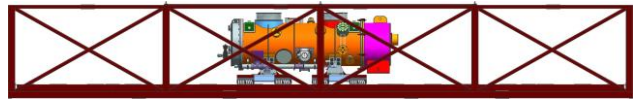

Figure 5: Modified LCLS-II Shipping Frame Supporting BCC Cryomodule

To determine the natural frequency of the cryomodule, a simplified shell and beam element body was created using the dimensions from the main cryomodule. The created model can be seen in Figure 6. The model was solved to find the first 100 mode frequencies of the assembly which resulted in a range of model frequencies between 9.3 to 76 Hz being found. In the range of resonant frequencies found, five were found to be bellow the 15 Hz limit and five more were within 5 Hz of the limit frequency. The range of frequencies found can be seen in Figure 7.

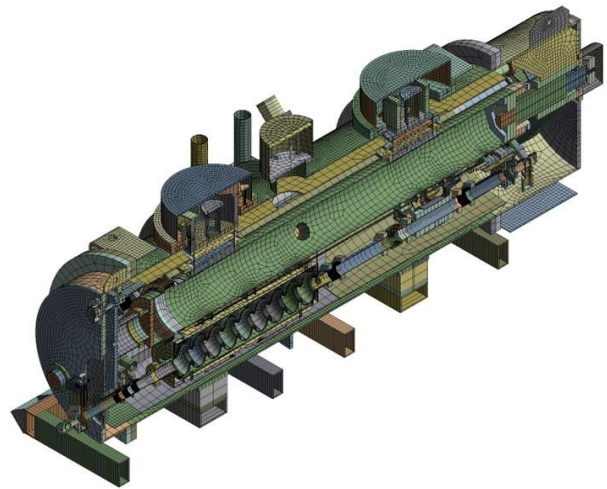

Figure 6: BCC Cryomodule Shell Body Model

Examining the resonant frequencies in the range, a majority of the frequencies were occurring in piping and wiring. From the simulations, some additional insulation and constraints were to the assembly. To further verify the assembly will survive the shipment, the resonant shapes were applied on the model and a static structural assessment was performed. Examining the deformed models at the ex-pected transportation accelerations, it was determined that the modal frequencies that could occur would not fail.

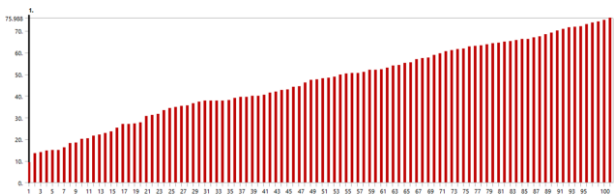

Figure 7: Resonant Frequencies of BCC Cryomodule

*STRUCTURAL ANALYSIS*

To incorporate all of the loading scenarios and account for the load paths through the assembly, three sets of Finite Element Analysis (FEA) simulation sets were created. The analyses looked at included load transfers from the endplates to the endcovers, a set of analysis for the Cryomodule Vessel, and a set of analysis for the Support Stands. Force and moment reactions were transferred between simulation sets to account for mechanical behavior between the complex loading conditions.

The analysis of the endplates to the endcover examined the behavior of the piping system lines terminate into an end plate and the load paths to the endcover. Due to the number of pressure lines terminating FEA was used to determine the force distribution across the end plate. The reaction forces into the endplate were modeled as input forces into a assessment of the endcover behavior. Combining all of the loading scenarios across the entire cryomodule, four load combinations were created for the endplate and endcover. An example of the first load combination can be seen in Figure 8.

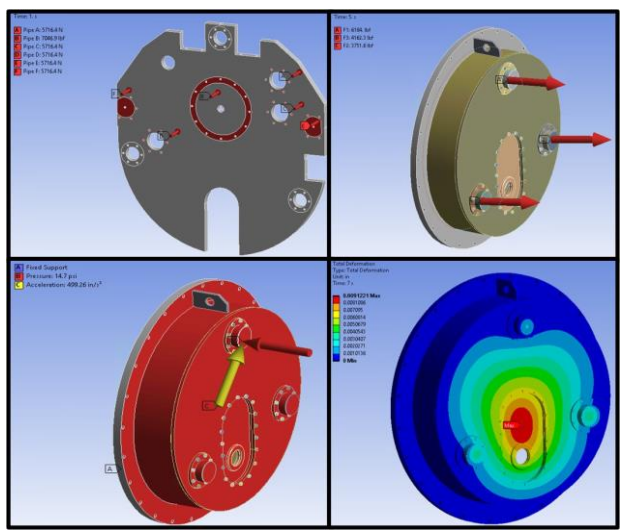

Figure 8: Endcover Boundary Conditions and Results: Endplate Boundary Conditions (Top-Left), Transferred Loads into Endcover (Top-Right), Endcover Boundary Conditions(Bottom-Left), Endcover Results (Bottom-Right)

Using the four load cases generated by the analysis of the endcover and endplate, six load cases were generated for the cryomodule addressing load combinations for operation, seismic loading, relief, and transportation scenarios. The analysis was performed to meet the Design by Analysis methods of ASME BPVC VIII Division 2 [6]. Calculations were also performed to meet the requirements for an ASME BPVC VIII Division 1 vessel [5]. The boundary conditions for the analysis can be seen in Figure 9 and the results for one of the load cases can be seen in Figure 10.

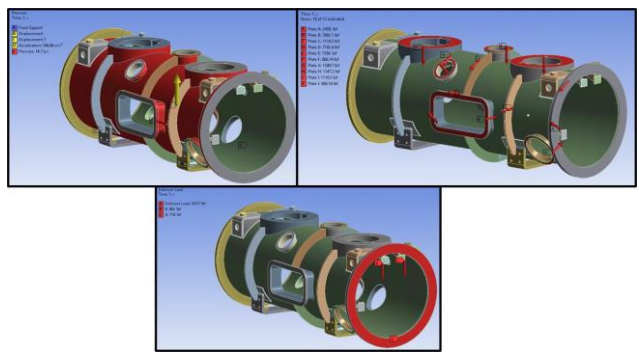

Figure 9: Cryomodule Boundary Conditions

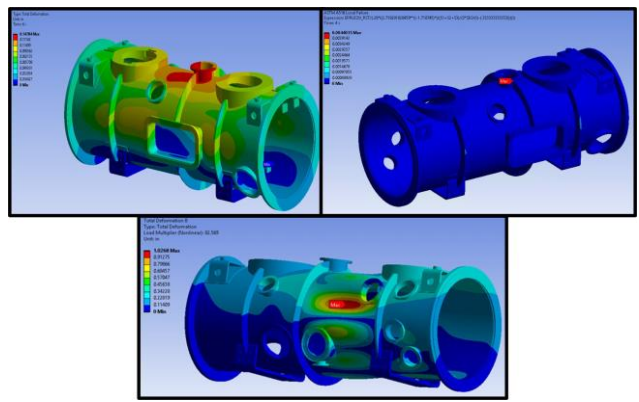

Figure 10: Cryomodule Results

Using reaction forces from the analysis of the cryomodule as loading forces on the stands, the stands were analyzed to determine if they would be able to support the cryomodule during all loading scenarios. To represent the stiffness of the cryostat, a cylinder was added to the same FEA model. Feedback from the initial calculations were used to reinforce the assembly. An example of the boundary conditions used on the updated stand design can be seen in Figure 11. The results of the analysis can be seen in Figure 12. The results of the manual calculations and FEA results indicate that the stands will be able to support the cryomodule under the projected loading scenarios.

**REVIEW STATUS**

The BCC Cryomodule is already past the Preliminary Design Review phase and will be soon going through the Final Design Review process. In preparation of these reviews, peer reviews within FNAL and SLAC have been performed. Reviews of the magnet design, Support Stands, the Pressure Vessel System, and the Transportation have been performed. Incorporating the feedback into the design process indicates that the design is sufficiently ready to enter the Final Design Review process.

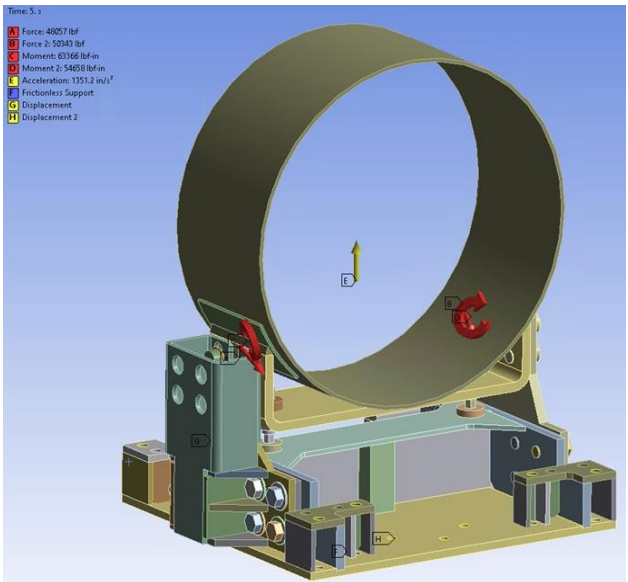

Figure 11: Cryomodule Stand Analysis Boundary Conditions

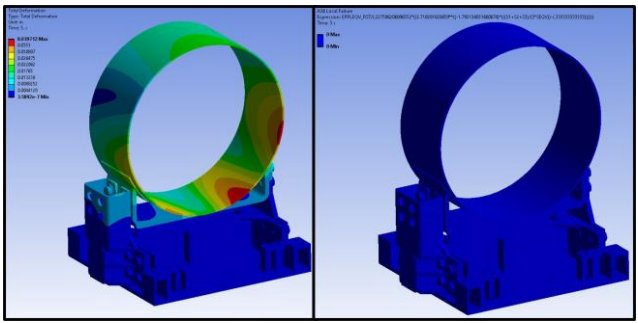

Figure 12: Cryomodule Stand Analysis Results

## ACKNOWLEDGEMENTS

This manuscript has been authored by Fermi Research Alliance, LLC under Contract No. DE-AC02-07CH11359 with the U.S. Department of Energy, Office of Science, Office of High Energy Physics.